\documentclass[aps,prb, twocolumn]{revtex4}
\usepackage{graphicx}
\usepackage{amsfonts}
\usepackage{amsmath}
\usepackage{amssymb}

\renewcommand{\Re}{\mathop{\rm Re}}
\renewcommand{\Im}{\mathop{\rm Im}}
\newcommand{\sign}{\mathop{\rm sign}}

\begin{document}

\title{Magnetic interference patterns in long disordered Josephson junctions}
\author{Beno\^it Crouzy}
\affiliation{Group AHEAD, Institute of Environmental Engineering, 
Ecole Polytechnique F\'ed\'erale de Lausanne, 1015 Lausanne, Switzerland}
\author{Dmitri A.\ Ivanov}
\affiliation{Institute for Theoretical Physics, ETH Zurich, 8093 Zurich, Switzerland}
\affiliation{Institute for Theoretical Physics, University of Zurich, 8057 Zurich, Switzerland}

\begin{abstract}
We study a diffusive superconductor--normal metal--superconductor
(SNS) junction in an external magnetic field. In the limit of a long
junction, we find that the form of the dependence of the Josephson current 
on the field and on the length of the junction depends on the
ratio between the junction width and the length associated with 
the magnetic field. A certain critical ratio between these two length scales
separates two different regimes. In narrow junctions, the critical current
exhibits a pure decay as a function of the junction length or of the
magnetic field. In wide junctions, the critical current exhibits
damped oscillations as a function of the same parameters. This
damped oscillating behavior differs from the Fraunhofer behavior
typical for short or tunnel junctions. In wide and long junctions, 
superconducting pair correlations and supercurrent are localized 
along the edges of the junction.
\end{abstract}


\maketitle

\section{Introduction}

Interplay between magnetism and superconductivity in proximity
structures results in rich physics.\cite{buzdin:05,bergeret:05} 
One of the remarkable effects
arising from magnetism in Josephson junctions is
the possibility of a reversal of the critical current ($\pi$ coupling).
It was theoretically 
predicted \cite{bulaevskii:77} and then experimentally 
observed \cite{pi-junction} in Josephson junctions with a ferromagnetic 
interlayer. In those junctions, $\pi$ coupling results from
the effective Zeeman field induced by the exchange interaction
in the ferromagnet.

In the present paper, we revisit
another way to reverse the critical current in a Josephson
junction by applying an external magnetic field. In this case,
the main role of the field is to provide a vector potential coupled
to the motion of electrons. This effect was studied in detail in the case 
of a thin (tunnel) junction. If a transverse magnetic field is applied to such a junction, 
then the Josephson coupling oscillates along the direction transverse to
the field, so that the total current exhibits  Fraunhofer-like oscillations
as a function of the magnetic flux through the junction.\cite{josephson:64,barone:82} 

It has been shown both theoretically\cite{bergeret:07:08} and
experimentally\cite{angers:08, chiodi:12} that the Fraunhofer interference
pattern does not appear in narrow and long junctions.
In particular, the authors of Ref.~\onlinecite{bergeret:07:08} 
have shown numerically how the damped oscillatory behavior (Fraunhofer like) characterizing 
wide and short junctions is replaced by a monotonic exponential decay in narrow diffusive
junctions. They have also identified the length scale over which the transition 
between the two regimes takes place. In the clean case, studies of similar SNS junctions 
in a magnetic field were performed in Ref.~\onlinecite{galaiko-gogadze-antsygina}
and, in a slightly different geometry, in Ref.~\onlinecite{mohammadkhani:08}. 

In this work, we revisit the problem of a diffusive SNS junction in
a magnetic field considered in Ref.~\onlinecite{bergeret:07:08} and show that further 
progress may be achieved in the limit of a long junction using analytical methods. 
The external magnetic field introduces a ``magnetic'' length scale
\begin{equation}
\xi_H=\sqrt\frac{\phi_0}{H}\, ,
\label{magnetic-length}
\end{equation} 
where $\phi_0=h/2e$ is the (superconducting) flux quantum and $H$ is
the external magnetic field. We confirm the findings
of Ref.~\onlinecite{bergeret:07:08} that the structure of the Josephson current
depends on the relation between the width of the junction
and $\xi_H$. For narrow junctions, the critical Josephson current decays 
exponentially as a function of the length of the junction (or, equivalently,
of the magnetic flux through the junction).  For wide junctions, the
critical Josephson current exhibits damped oscillating behavior
(reversing sign), as a function of either the length of the junction
or of the total magnetic flux through it. The period of these
oscillations is the same as in the Fraunhofer limit but the decay
is exponential instead of algebraic. In this wide-junction regime, the
superconducting correlations and the supercurrent are localized 
near the edges of the junction. We find analytically the
transition point (the critical width of the junction) where the 
monotonic decay of the critical current switches over to damped oscillations.

The paper is organized as follows. In Section \ref{section:junction} 
we describe the model of a diffusive SNS junction in a transverse magnetic field.
We compute the superconducting Green function in Section \ref{section:SN} and 
the Josephson current in Section \ref{section:current}.
In Section \ref{section:summary} we summarize our results. Several
technical details are relegated to the Appendices.

\section{SNS junction in a transverse magnetic field}
\label{section:junction}

We consider a SNS junction in a transverse magnetic field
in a quasi-two-dimensional geometry.
A coordinate system is introduced in such a way that the $x$ axis is
directed along the junction, the field is applied along the $z$
axis, and the SN interfaces are parallel to the $yz$ plane (Fig.~1). 
The SN interfaces correspond to the coordinates $x=0,L_x$.  
The origin of the $y$ axis is chosen in the middle of one of the SN interfaces 
and we denote the width of the junction by $L_y$ (so that the edges of the junction
correspond to $y=\pm L_y/2$).  The system is translationally invariant 
along the $z$ direction. We assume a uniform magnetic field $\mathbf{H}$ directed
along the $z$ axis and neglect the screening of the magnetic field by the
Josephson currents (which assumes that the dimensions
of the junction in the $y$ and $z$ directions are much shorter than the 
Josephson penetration length).\cite{bergeret:07:08} 

For simplicity, we further assume that the London length is short compared 
to other length scales in the problem and neglect the penetration of the magnetic 
field in the superconducting electrodes. Then the vector potential 
of the field may be chosen as $\mathbf{A}=-yH\mathbf{e}_x$, with the phase of the 
superconducting order parameter (denoted $\pm\chi$ below) constant
along the SN interface in each of the superconducting leads (see Appendix~\ref{appendix:gauge}).
This choice of the vector potential was used previously in 
Ref.~\onlinecite{bergeret:07:08,belzig:96} and is exactly gauge equivalent
to the more conventional gauge with $\mathbf{A}$ parallel to the SN
interfaces considered in Ref.~\onlinecite{galaiko-gogadze-antsygina}.

We further assume that the normal layer is strongly disordered and the 
motion of electrons is diffusive. In this regime, the quasiclassical 
Green functions (averaged over the fast Fermi oscillations and the momentum directions) 
are given by the solutions to the Usadel equations:\cite{usadel:70,kopnin:01}
\begin{equation}
\hbar{D}\hat{\mathbf{\nabla}} \left( \check{g}\hat{\mathbf{\nabla}} 
\check{g}\right) - \omega
\left[ \hat{\tau}_{3},\check{g}\right] =0\, , 
\qquad
\check{g}^{2}={\check{1}}\, .
\label{usadel}
\end{equation}
Here $D$ is the diffusion constant, 
\begin{equation}
\omega =\left( 2n+1\right) \pi T
\label{Matsubara}
\end{equation} 
is the Matsubara frequency, and $\hat{\tau}_{3}$ is a Pauli matrix
in the particle-hole (Nambu) space. The Green function may be parametrized as
\begin{equation}
\check{g}=\left( \begin{array}{cc} G & F \\
F^{\dagger} & -G \end{array}\right)
\end{equation}
with a real $G$ and complex conjugate $F$ and $F^\dagger$
(our definition of $F$ and $F^\dagger$ differs from that
of Ref.~\onlinecite{kopnin:01} by a factor of $i$).
The gradient operator $\hat{\mathbf{\nabla}}$ in Eq.~(\ref{usadel})
includes the vector potential:
\begin{equation}\label{pb3} 
\hat{\mathbf{\nabla}}= \mathbf{\nabla} 
- \frac{ie}{\hbar}\mathbf{A} [\hat{\tau}_3, \,\cdot\; ]\, .
\end{equation}
We neglect the Zeeman splitting, which, in the case
of the transverse magnetic field, has a typically much smaller effect 
than the vector-potential term.

The Usadel equations (\ref{usadel}) are complemented by the boundary
conditions at the edges of the junction
\begin{eqnarray}
\nabla_y \check{g}(x,y=\pm L_y /2 ) = 0
\label{bc3}
\end{eqnarray}
and by suitable boundary conditions at the SN interfaces ($x=0$ and $x=L_x$),
depending on the transparency of the interface, the levels of disorder
in the N and S regions, and the ratio of the Matsubara frequency to the
superconducting gap.\cite{boundary-conditions} For simplicity, we assume
that the superconducting gap is the largest energy scale in the problem
and that the SN interfaces are perfectly transparent: then the boundary
conditions take the simplest ``rigid'' form
\begin{equation}
\check{g}(x=\{ 0, L_x\} ,y) = \left( \begin{array}{cc}
0 & e^{\pm i\chi} \\ e^{\mp i \chi} & 0 
\end{array} \right)\, .
\label{rigid-bc}
\end{equation}
The boundary conditions at the SN interfaces will not
be important for our result on the transition between the
purely decaying and oscillating-decaying regimes, but will
play a role in estimating the pre-exponential factor in 
Eq.~(\ref{current3}) below.

There are several energy scales in the problem. We assume that the
gap in the superconducting terminals is the largest energy scale.
In particular, it is much larger than the energy scale associated
with the magnetic length $\xi_H$ (the ``magnetic Thouless energy'')
defined as
\begin{equation}
E_H=\frac{\hbar D}{\xi_H^2}=\frac{DeH}{\pi}
\label{EH}
\end{equation}
This energy scale is, in turn, assumed to be much higher than the 
temperature $T$. These assumptions agree with the high-field
regime of the model considered in Ref.~\onlinecite{bergeret:07:08}
and of the experiments in Refs.~\onlinecite{angers:08,chiodi:12}.
The conventional Thouless energy 
\begin{equation}
E_{\rm Th}= \frac{\hbar D}{L_x^2}
\label{Thouless}
\end{equation}
(which is smaller than $E_H$)
plays minor role in our long-junction (or large-field) limit, since the
decay of the proximity correlations is mainly determined by
the magnetic length $\xi_H$.

\begin{figure}
\includegraphics[width=0.35\textwidth]{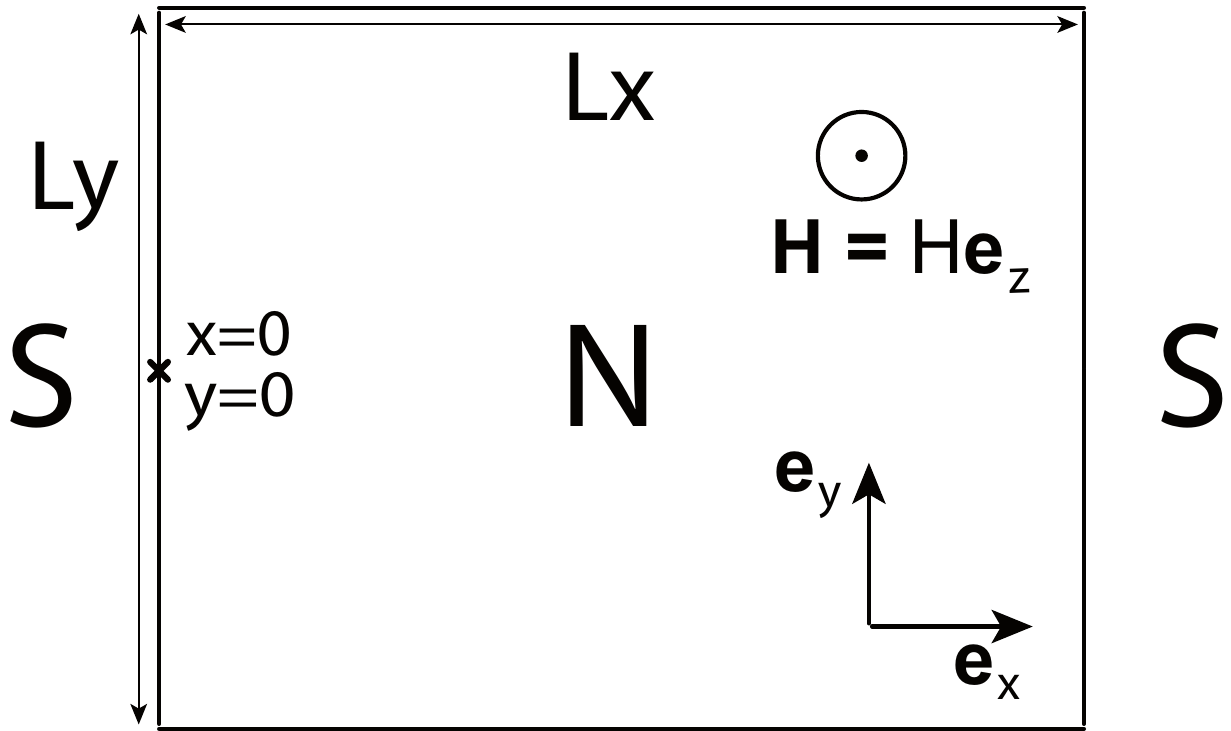} 
\caption{SNS junction in a transverse magnetic field.}
\label{setup}
\end{figure}

The current density can be calculated from $F(x,y)$ as\cite{kopnin:01}
\begin{equation}
\mathbf{J}=2\pi i e \nu DT\,\sum_{n=0 }^{\infty }\left[ F^{\dagger
}\mathbf{\nabla }F-F\mathbf{\nabla }F^{\dagger
}-\frac{4ie\mathbf{A}}{\hbar}FF^{\dagger}\right]\, ,
\label{current}
\end{equation}
where $\nu$ is the density of states in the normal metal phase (per one
spin projection). The symmetry of translation along the $z$ direction implies 
that the current remains in the $xy$ plane. The sum is taken over the Matsubara 
frequencies (\ref{Matsubara}).

\section{Decay of superconducting correlations in a long junction}
\label{section:SN}

For a sufficiently long junction (a specific condition will be determined below),
the anomalous correlations $F$ in the middle
of the junction are exponentially weak. Therefore, to the leading order, we
may approximate the solution to the SNS-junction problem as a superposition
of two solutions to the semi-infinite SN problem\cite{Zaikin-Zharkov:81,ivanov:02}
\begin{equation}
F(x,y) \approx F_{\infty}(x,y) e^{i\chi}+F_{\infty}(L_x{-}x,-y) e^{-i\chi}
\label{weak-coupling}
\end{equation}
where $F_{\infty}(x,y)$ is the off-diagonal matrix element
of the (full nonlinear) solution $\check{g}_\infty$ for the SN problem with 
the semi-infinite normal layer. It obeys the same equations (\ref{usadel}) with the
same boundary conditions at $x=0$ and at $y=\pm L_y/2$ and  with the boundary at $x=L_x$ 
replaced by $\check{g}_{\infty}(x\to\infty ,y)=\sign(\omega) \hat{\tau}_3$. 
This weak-coupling regime, via
the expression (\ref{current}), implies the sinusoidal current-phase 
relation\cite{golubov:04} 
\begin{equation}
J_\mathrm{tot}=I_{c}\sin 2\chi \, ,  \label{cpr}
\end{equation}
where $J_\mathrm{tot}$ is the total Josephson current 
(integrated over the $y$ and $z$ directions).

The problem now reduces to finding the mode of $F_\infty(x,y)$ with the slowest
decay (along the $x$ direction) in the auxiliary problem of a semi-infinite 
infinite SN junction. At sufficiently large $x$, the proximity effect is
weak, and $F_\infty(x,y)$ obeys the linearized version of the Usadel equations
(\ref{usadel}) [obtained by assuming $G\approx\sign(\omega)$ and $|F|\ll 1$]. For the sake of convenience, 
in all the equations below we rescale both  coordinates $x$ and $y$ in the units 
of the magnetic length (\ref{magnetic-length}). Then the linearized Usadel
equation takes the form\cite{bergeret:07:08}
\begin{equation}
\left[ \left( \nabla_x + 2 \pi i y\right)^2 + \nabla_y^2 - 
\frac{2|\omega|}{E_H} \right] F_\infty(x,y) =0\, .
\label{lin}
\end{equation}

The mode with the slowest decay may be written as
\begin{equation}
F_\infty(x,y) \sim e^{-\kappa x}\, \psi(y)\, , \qquad x \to \infty\, ,
\label{F-asymptotic}
\end{equation}
where $\kappa$ is the inverse decay length. By substituting this expression
into the linearized Usadel equation, we find that $\psi(y)$ must be a zero
mode of the linear operator
\begin{equation}
\left[ \nabla_y^2 +  \left( \kappa - 2 \pi i y\right)^2 - \frac{2|\omega|}{E_H} \right] \psi(y) = 0
\label{zero-mode}
\end{equation}
with the boundary conditions $\psi'(\pm L_y/2)=0$. The mode with the
slowest decay may be found as the zero mode (\ref{zero-mode}) with the
smallest positive real part of $\kappa$.

\begin{figure}
\includegraphics[width=0.4\textwidth]{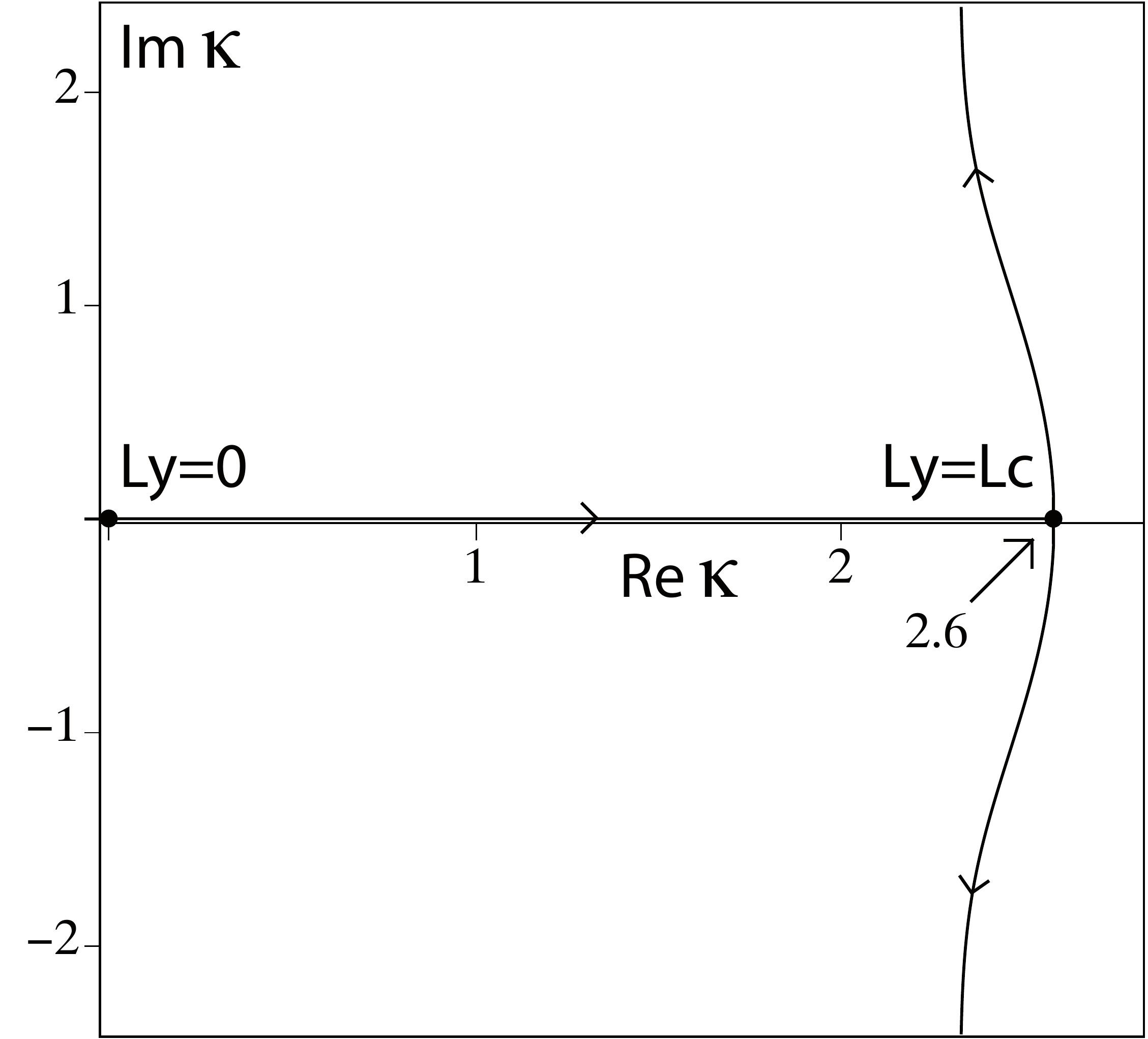} 
\caption{Effective inverse decay length $\kappa$ (in the units of $\xi_H^{-1}$). 
A purely real $\kappa$ indicates a monotonic decay. The transition to a 
damped oscillating regime occurs at $L_y=L_c\approx0.82$ (in the units of $\xi_H$)
and corresponds to $\kappa\approx 2.6$.}
\label{fig:effective}
\end{figure}

One can verify that the real part of $\kappa$ given by the zero-mode condition 
(\ref{zero-mode}) increases with increasing $|\omega|$. Therefore, under
the low-temperature assumption $T\ll E_H$,
the main contribution to $F_\infty(x,y)$ at large $x$ comes
from Matsubara frequencies $\omega \ll E_H$. For these frequencies,
we may neglect the $\omega$ term in Eq.~(\ref{zero-mode}). 
The general solution to the second-order differential equation 
(\ref{zero-mode}) at $\omega=0$
can be written in terms of a linear combination of two modified
Bessel functions,\cite{abramovitz:72}
\begin{multline}
\psi(y) = \sqrt{i\kappa+2\pi y} \Bigg( C_1 \, I_{1/4}\left[\frac{(i\kappa+2\pi y)^2}{4\pi}\right] \\
+ C_2 \, K_{1/4}\left[\frac{(i\kappa+2\pi y)^2}{4\pi}\right] \Bigg)\, .
\label{exact}
\end{multline}
The boundary conditions at $y= \pm L_y/2$ fix the ratio $C_1 / C_2$ and
limit the possible values of $\kappa$ to a discrete set.

As a result, we obtain the values of $\kappa$ (in the units of $\xi_H^{-1}$)
as a function of $L_y$ (in the units of $\xi_H$). The trajectory
of $\kappa$ in the complex plane, as the width of the 
junction $L_y$ varies from zero to infinity, is plotted
in Fig.~\ref{fig:effective}. 

In the limit $L_y \to 0$, the $2\pi i y$ term in
the operator (\ref{zero-mode}) may be neglected, and the spectrum
is composed of the non-degenerate eigenvalues $\kappa= n\pi/L_y$.
The ``leading'' eigenvalue with the smallest real part is thus 
zero at $L_y \to 0$. At a small finite $L_y$, 
the leading eigenvalue $\kappa$ also becomes finite, but remains purely real.
This follows from the combined symmetry of the complex conjugation
and the reflection $y \mapsto -y$, which relates the eigenvalues
$\kappa$ and $\kappa^*$. Since the leading eigenvalue is nondegenerate in the
limit $L_y \to 0 $, by continuity it must remain purely imaginary
for sufficiently small $L_y$.\cite{note:skvor}

At larger $L_y$, two real eigenvalues may collide and bifurcate
to a complex-conjugate pair. This happens at $L_y=L_{c}\approx0.82$ 
(see Fig.~\ref{fig:effective}). For $L_y>L_c$, 
the contributions of the two modes (corresponding
to the complex conjugate pair $\kappa$ and $\kappa^*$) decay with the same
rate (given by the real part of $\kappa$) and must be added together. 
In the discussion of the wide-junction limit (Section \ref{subsec:wide}), 
we will show that those modes, in the limit $L_y\gg 1$, correspond 
to solutions localized at the two edges of the junction $y=\pm L_y /2 $. 
The critical length $L_c$ separates the regime where the superconducting anomalous
Green function $F(x,y)$ decays along the $x$ direction without oscillations
(narrow junction, purely real $\kappa$) and the regime where the decay
of the Green function is damped oscillatory (wide junction, 
a pair of complex conjugate values of $\kappa$).

\subsection{Narrow-junction limit}
\label{subsec:narrow}

For $L_y\ll 1$ (in the units of $\xi_H$) 
we expand the exact solution (\ref{exact}) in powers
of $L_y$ and find the wave number $\kappa$ solving the
equation for the boundary conditions at
$y=L_y/2$. This yields the expansion
\begin{equation}
\kappa=\frac{\pi}{\sqrt{3}}\, L_y
\left(
1 +\frac{4 \pi^2}{63} L_y^4
+\frac{932 \pi^4}{218295} L_y^8 + \ldots \right)\, .
\label{narrow}
\end{equation}

To the lowest order in $L_y$, the solution to the Usadel
equation does not depend on $y$. In this limit, one can simply
average the $y^2$ term in the Usadel equation (\ref{lin}) and
arrive at a pair-breaking term \cite{bergeret:07:08,angers:08,crouzy:05}
\begin{equation}
\frac{\hbar{D}}{2}\nabla_{x}^2F(x)=\left(|\omega|+2\Gamma \right)F(x)
\label{usadelsf}
\end{equation}
with
\begin{equation}
\Gamma= \frac{\hbar D \kappa^2}{4}
=\frac{De^2H^2L_y^2}{12\hbar}\, .
\label{spin-flip}
\end{equation}
This result obviously reproduces the first term in Eq.~(\ref{narrow}).

\begin{figure}
\includegraphics[width=0.4\textwidth]{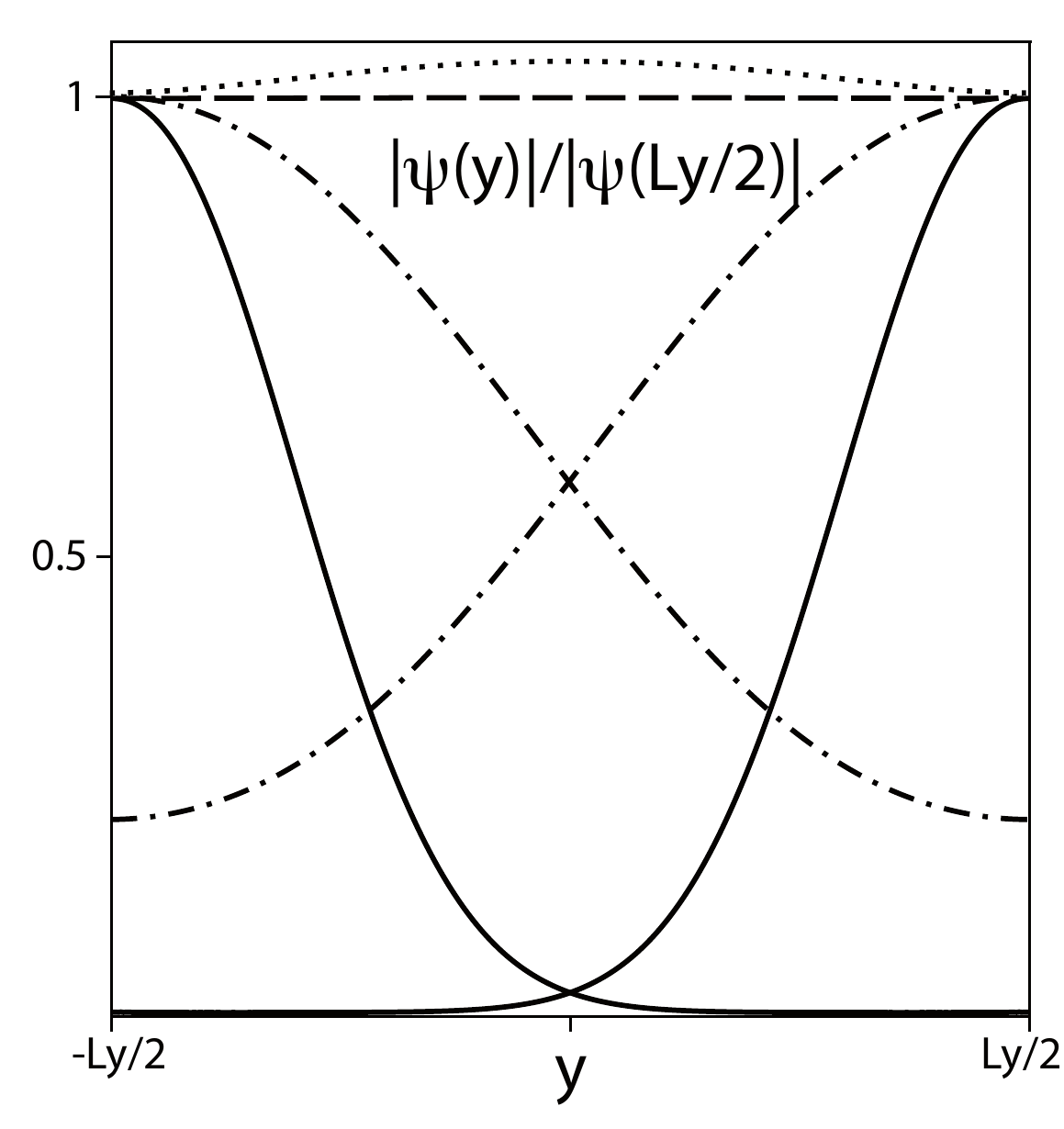} 
\caption{Superconducting pair correlations $\left|\psi\right|$ normalized to their value at the border of the junction for $L_y=0.25\xi_{H}$ (dash), $0.75\xi_{H}$ (dot), $\xi_{H}$ (dash dot) and $2.5\xi_{H}$ (solid).}
\label{fig:zero-modes}
\end{figure}

\subsection{Wide-junction limit}
\label{subsec:wide}

In the limit of a wide junction $L_y \gg 1$ (as usual, in the units of $\xi_H$), 
the solution is determined by the two complex conjugate wave vectors 
$\kappa$ and $\kappa^*$. We show below that
the asymptotic behavior of $\kappa$ (in the units of $\xi_H^{-1}$) is 
\begin{equation}
\kappa \approx i \pi L_y + \kappa_{\mathrm{res}}\, .
\label{kappa-res-1}
\end{equation}

Indeed, in the wide-junction limit, each of the two zero modes (\ref{exact})
is  localized near one of the two edges of the junction and
decays quasiclassically towards the other edge. The solution localized 
near $y=L_y/2$ should therefore have the quasiclassical wave vector 
in the operator (\ref{zero-mode}) vanishing in that region, which 
immediately gives the leading asymptotics $\kappa\approx i \pi L_y$
(the solution localized at the opposite edge has $\kappa \approx -i \pi L_y$).

To get the subleading term $\kappa_{\mathrm{res}}$, we consider one of those zero modes
(say, the one localized near $y=L_y/2$). This zero mode decays 
quasiclassically towards the opposite edge of the junction, and with
an exponential precision we can replace the boundary condition at
$y=-L_y/2$ by the decaying condition at infinity, $\psi(y\to-\infty)=0$.
This selects a solution from (\ref{exact}) of the form
\begin{equation}
\psi \propto \sqrt{i\kappa + 2\pi y}\, K_{1/4}\left[\frac{(i\kappa+2\pi y)^2}{4\pi}\right]\, .
\label{bessel-solution-2}
\end{equation}
Imposing now the boundary condition $\psi'(L_y/2)=0$ implies an
equation on $\kappa_{\mathrm{res}}$:
\begin{equation}
K_{3/4}\left[\frac{(-i \kappa_{\mathrm{res}})^2}{4\pi}\right] = 0\, .
\label{k-res-equation}
\end{equation}
A numerical solution to this equation gives\cite{note:Macdonald}
\begin{equation}
\kappa_{\mathrm{res}} \approx 2.32 - 1.68 i \, .
\label{kappa-res-2}
\end{equation}

We illustrate our calculation in Fig.~\ref{fig:zero-modes}, where
we plot the zero modes below and above the transition.
Below the transition (for $L_y<L_c$), the solution is nondegenerate
and symmetric, while above the transition ($L_y>L_c$) the two zero
modes are pushed towards the edges of the junction. The characteristic
size of the region near the edge where the proximity correlations
are localized is of the order one [from the solution (\ref{bessel-solution-2})],
i.e., $\xi_H$ in the physical units.

\section{Josephson current}
\label{section:current}

In Section~\ref{section:SN}, we have calculated the wave vectors $\kappa$
of decay of the superconducting correlations in a long SN junction. As
we shall see below, the same wave vector $\kappa$ describes the
dependence of the Josephson current on the length $L_x$ of a SNS junction.
In particular, the $L_x$ dependence of the Josephson current exhibits
two types of behavior: purely decaying (for $L_y<L_c$) 
and decaying with oscillations (for $L_y>L_c$).
This result is consistent with the previous numerical works,\cite{bergeret:07:08}
which identified these two regimes.

To calculate the Josephson current, we substitute the asymptotic
behavior (\ref{F-asymptotic}) of $F_\infty(x,y)$ (possibly containing
two terms in case of a complex $\kappa$), applicable in the middle
of the junction, into Eq.~(\ref{current}). It produces the sinusoidal
current-phase relation (\ref{cpr}). 

In the ``pure decay''
phase ($L_y<L_c$, real $\kappa$), the critical current $I_c$ is
given by
\begin{equation}
I_c = 8\pi e \nu D T L_z \sum_{n=0}^\infty
\left[ \int_{-L_y/2}^{L_y/2} (\kappa - 2\pi i y) \psi^2(y) dy \right]
e^{- \kappa L_x} \, .
\label{current1}
\end{equation}
where $L_z$ is the dimension of the junction along the $z$ direction.
Note that this expression is real [since $\psi(y) = \psi^*(-y)$ in
this regime] and positive (one checks this numerically).

In the regime of ``decaying oscillations'' ($L_y > L_c$, complex $\kappa$), the
anomalous Green function $F_{\infty}(x,y)$ contains contributions
from two zero modes:
\begin{equation}
F_{\infty}(x,y) = \psi(y) e^{-\kappa x} + \psi^*(-y) e^{-\kappa^* x} \, ,
\end{equation}
where $\psi(y) \ne \psi^*(-y)$ are the two leading
zero modes (\ref{zero-mode}).
Integrating the critical current along the $y$ and $z$ directions,
one finds
\begin{equation}
I_c = 8\pi e \nu D T L_z \sum_{n=0}^\infty
\, \Re 
 \int_{-L_y/2}^{L_y/2} (\kappa - 2\pi i y) \psi^2(y) dy \, e^{-\kappa L_x}
\, .
\label{current2}
\end{equation}
Note that in the case of a wide junction, $L_y \gg \xi_H$, the
localization of the superconducting pair correlations at the edges
of the junction results in the localization of the current 
in the same region.

In Eqs.\ (\ref{current1}) and (\ref{current2}), both $\kappa$ and
$\psi(y)$ depend on the index $n$ of the Matsubara frequency (\ref{Matsubara}).
Since $\Re\kappa$ grows with $\omega$, only
terms with $\omega \ll E_H$ contribute in the large-$L_x$ limit. 
Therefore, the leading exponential dependence of $I_c$ on $L_x$
is given by
\begin{equation}
I_c \sim \Re \, I_c^{(0)} \exp [- \kappa L_x]\, ,
\label{current3}
\end{equation}
where $\kappa$ is now taken at $\omega=0$ (plotted in Fig.~\ref{fig:effective}).
The above formula is equally applicable in the ``pure decay'' phase (with
real $\kappa$ and $I_c^{(0)}$) and in the ``decaying oscillations'' phase
(with complex $\kappa$ and $I_c^{(0)}$).
$\Re \kappa$ and $\Im \kappa$ describe the rates of decay 
and oscillations of the critical current as a function of $L_x$, respectively.

An accurate calculation of $I_c^{(0)}$ would require knowing the normalization
of $\psi(y)$ (which depends on the specific boundary conditions at the NS interface
and may require solving the nonlinear Usadel equations near the interface) and
the $\omega$ dependence of $\kappa$. We do not perform such a detailed
calculation, but only give order-of-magnitude estimates, under the assumption
of a perfectly transparent SN interfaces with rigid boundary conditions 
(\ref{rigid-bc}) for the Usadel equations (see Appendix \ref{appendix:current} for details).

In the low-temperature regime (below a certain temperature scale $T_*$), 
many Matsubara frequencies contribute to the total current (\ref{current1})
or (\ref{current2}), while at temperatures above $T_*$ only the
lowest Matsubara frequency is relevant. A simple estimate 
(see Appendix \ref{appendix:current}) gives
\begin{equation}
T_* \sim \begin{cases}
\frac{\phi}{\phi_0} E_{\rm Th} = \frac{L_y}{L_x} E_H\, , & L_y \ll \xi_H \, , \\
\sqrt{E_H E_{\rm Th}} = \frac{L_x}{\xi_H} E_{\rm Th}\, , & L_y \gg \xi_H\, ,
\end{cases}
\label{T-star}
\end{equation}
where $E_{\rm Th}$ is the Thouless energy defined in Eq.~(\ref{Thouless})
and $\phi = H L_x L_y$ is the total flux through the junction.
Note that in these estimates and in the estimates below we omit numerical 
coefficients which may reach an order of magnitude.\cite{Zaikin-Zharkov:81,Dubos:01}

For narrow junctions, $L_y \ll \xi_H$, we find (see Appendix \ref{appendix:current})
\begin{equation}
I_c^{(0)} \sim \begin{cases}
I_0 \left( \frac{\phi}{\phi_0} \right)^2\, , & T \ll T_*\, , \\
I_0 \frac{T}{E_{\rm Th}} \left( \frac{\phi}{\phi_0} \right)\, , & T \gg T_*\, ,
\end{cases}
\label{current-narrow-1}
\end{equation}
where $I_0$ is the critical current through the same junction
at low temperature ($T \ll E_{\rm Th}$) {\it without} magnetic field.

For wide junctions, $L_y \gg \xi_H$, the critical current may be
estimated (see Appendix \ref{appendix:current}) as
\begin{equation}
I_c^{(0)} \sim \begin{cases}
I_0 \sqrt{\frac{\phi}{\phi_0}} \left( \frac{L_x}{L_y} \right)^{3/2}\, , & T \ll T_*\, , \\
I_0 \frac{T}{E_{\rm Th}} \left( \frac{L_x}{L_y} \right)\, , & T \gg T_*.
\end{cases}
\label{current-wide-1}
\end{equation}

\begin{figure}
\includegraphics[width=0.4\textwidth]{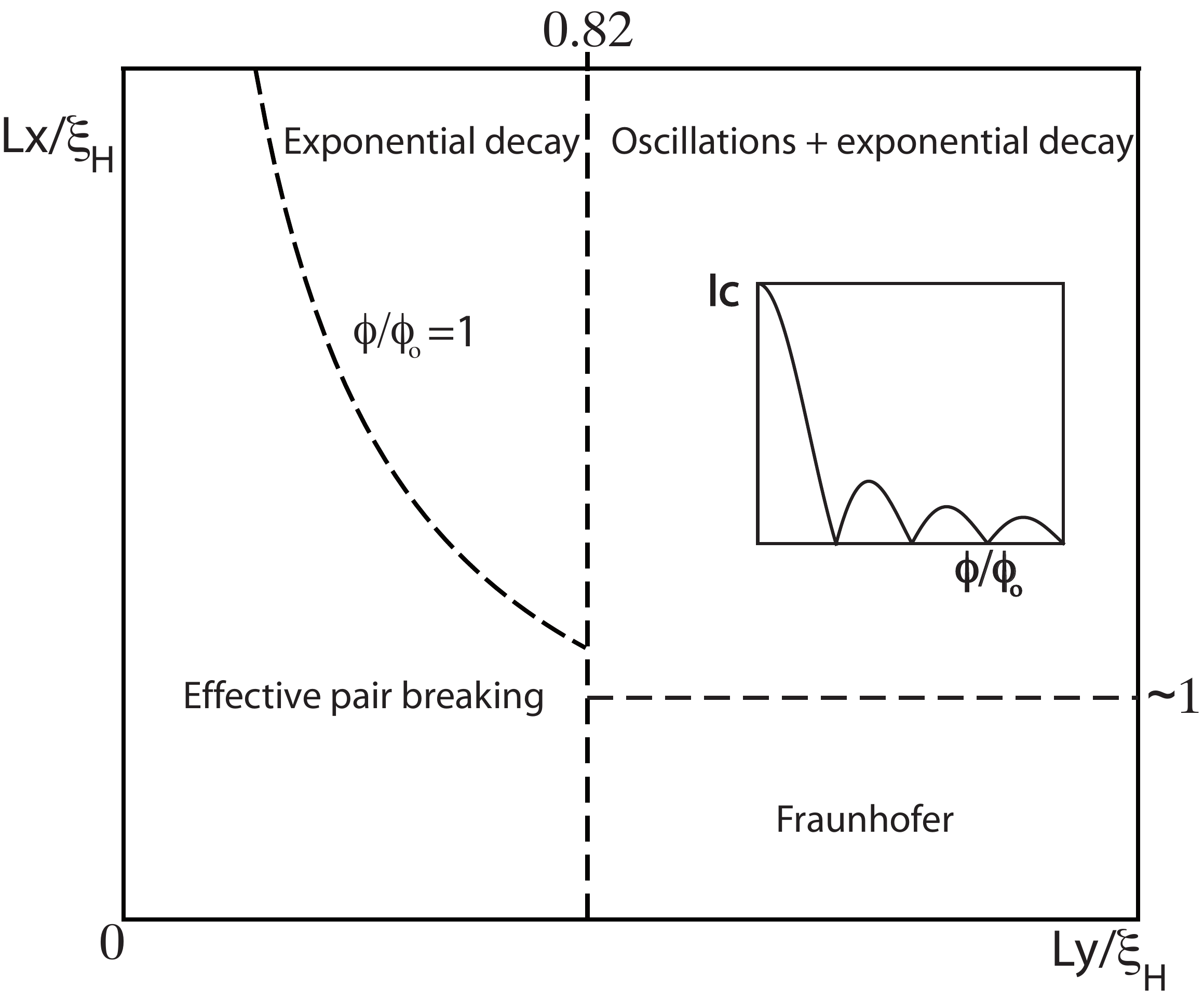} 
\caption{The phase diagram of the junction in a magnetic field. 
In the left region ($L_y/\xi_H < 0.82$), the critical current monotonically decays
as a function of the field. In the right region ($L_y/\xi_H > 0.82$), the current
oscillates and decays as a function of the field.}
\label{fig:phase}
\end{figure}

We sketch the phase diagram of the junction in Fig.~\ref{fig:phase}
in the coordinates $L_x$ and $L_y$. 

In the region $L_y<L_c$, there is
no sign reversal of the Josephson coupling (no $\pi$ phase). The condition
of the applicability of our result (\ref{current3}) is $\kappa L_x \gg 1$,
which translates in this region to $\phi/\phi_0 \gg 1$. This is
exactly the condition of a ``sufficiently long'' junction which
justifies the weak-coupling approximation (\ref{weak-coupling}).
At weaker fields (or for shorter junctions), $\phi/\phi_0 \le 1$,
the expression (\ref{current3}) is not accurate, but the pair-breaking
approximation (\ref{usadelsf})--(\ref{spin-flip}) remains valid, 
provided $L_y \ll \xi_H$. 

In the region $L_y > L_c$, there are
sign reversals of the Josephson coupling, and the oscillating
exponential decay (\ref{current3}) is applicable as long as
$L_x \gg \xi_H$ (since $\Re \kappa$ is of the order one in this
phase). For shorter junctions, $L_x \ll \xi_H$, the exponential 
formula (\ref{current3}) crosses over to the conventional
Fraunhofer interference pattern \cite{josephson:64,barone:82} (see Appendix
\ref{appendix:Fraunhofer} for details).

In experiments, one usually varies the external field for a 
junction of fixed dimensions. This corresponds to scanning
the phase diagram in Fig.~\ref{fig:phase} along a diagonal
with a fixed ratio $L_x/L_y$. Then, depending on this ratio,
one may observe a crossover from the pure-decay (for $L_y \ll L_x$) 
or Fraunhofer (for $L_y \gg L_x$)
regime to the exponential oscillating regime as the field
increases. 

For junctions with $L_y \ll L_x$, the low-field regime
($\xi_H \gg L_y$) corresponds to the ``effective spin flip'' and
``exponential-decay'' regions of the phase diagram in Fig.~\ref{fig:phase}.
There, the field dependence of the critical current
is given [by combining Eqs.\ (\ref{current3}) and (\ref{current-narrow-1})] by
\begin{equation}
I_c \propto \left(\frac{\phi}{\phi_0} \right)^{\alpha_1} 
\exp \left( - \frac{\pi}{\sqrt{3}}\, \frac{\phi}{\phi_0} \right)\, ,
\label{current-field-narrow}
\end{equation}
where $\alpha_1=2$ for $T \ll T_*$ and $\alpha_1=1$ for $T \gg T_*$ 
(note that $T_*$ itself depends on the magnetic field). The above
expression is only applicable in the $\phi \gg \phi_0$ limit.
Note that our $T \gg T_*$ result agrees with the $\phi \gg \phi_0$
limit of the more general expression
\begin{equation}
I_c \propto \frac{ \frac{\pi}{\sqrt{3}}\, \frac{\phi}{\phi_0}}{
\sinh \left(  \frac{\pi}{\sqrt{3}}\, \frac{\phi}{\phi_0} \right)}
\label{current-field-Montambaux}
\end{equation}
derived in Ref.~\onlinecite{montambaux:07} under the assumption
of the sinusoidal current-phase relation and with the result
of Ref.~\onlinecite{hammer:07} in the context of a spin-flip scattering.

Upon increasing the magnetic field, at $\xi_H \ll L_y$, the same junction crosses 
over to the ``oscillations and decay'' region of the phase diagram in 
Fig.~\ref{fig:phase}. There, the critical current 
exhibits the decaying-oscillating behavior [found by 
combining Eqs.\  (\ref{current3}) and (\ref{current-wide-1})]:
\begin{equation}
I_c \propto \left(\frac{\phi}{\phi_0} \right)^{\alpha_2} 
\exp \left[ - 2.32\, \frac{L_x}{\xi_H} \right]
\sin \left[ \frac{\pi \phi}{\phi_0} -1.68 \frac{L_x}{\xi_H}
+ \varphi_0 \right]\, ,
\label{current-field-wide}
\end{equation}
where $\alpha_2=1/2$ for $T \ll T_*$ and $\alpha_2=0$ for $T \gg T_*$.
Here $\varphi_0$ is a phase shift, which we do not compute
in this work. The asymptotic expression (\ref{current-field-wide})
is applicable at $L_x,L_y \gg \xi_H$, which implies, in particular,
$\phi \gg \phi_0$. Note that while both expressions (\ref{current-field-narrow}) and
(\ref{current-field-wide}) decay exponentially with increasing the
field, the expression in the exponent of Eq.~(\ref{current-field-narrow})
is proportional to $H$, while that in the exponent of 
Eq.~(\ref{current-field-wide}) only to $\sqrt{H}$.

For junctions with $L_y \ll L_x$, the low-field regime ($\xi_H \gg L_x$) belongs to the
Fraunhofer region of the phase diagram in Fig.~\ref{fig:phase}.
In this region, a simple analysis of our model 
produces the usual Fraunhofer pattern of the current-field dependence \cite{josephson:64,barone:82}
with $1/\phi$ decay and oscillations with the period  $\phi_0$  (see Appendix~\ref{appendix:Fraunhofer}).
In the same junction, with the increase of the field ($\xi_H \ll L_x$), 
this Fraunhofer pattern crosses over to the oscillating-decay regime (\ref{current-field-wide}).
In this crossover, the period of oscillations (as a function of the field) remains the same, 
only the $1/\phi$ rate of decay gets replaced by an exponential.

\section{Summary and discussion}
\label{section:summary}

To summarize, we consider a long diffusive SNS junction in an external
magnetic field $\mathbf{H}$. We show that depending on the width of
the junction relative to the magnetic length
$\xi_H=\sqrt{\phi_0 / H}$ two different regimes can be
observed. For narrow junctions the anomalous Green function $F$ decays
monotonically along the junction while for wide junctions
exponentially damped oscillations are present. We find that the
transition between the two regimes occurs at the width
$L_c\approx0.82\xi_H$. These different behaviors of
the proximity correlations translate into two
types of the dependence of the Josephson critical current
on the magnetic flux through the junction: a monotonic decay
for narrow junctions and an oscillating decay for wide junctions.
We also show that for wide junctions, both the superconducting pair 
correlations and the current are concentrated 
in a small region of size $\xi_H$ near the edges of the junction.

The main finding of the present work, in comparison with previous
studies of this problem, is the identification of the
damped-oscillating phase for wide and long junctions. This phase
may be understood as a ``crossover'' between the Fraunhofer and
the purely damped regimes: the period of
oscillations is the same as in the Fraunhofer interference pattern,
while the exponentially decaying factor resembles the damped phase.

Conceptually, the transition between the two asymptotic regimes
for long junctions in our problem is similar to the transition
between the two regimes in superconductor--ferromagnet--superconductor
junctions with domains studied in Ref.~\onlinecite{crouzy:07}.
In both systems, the
transition between the purely damped and damped-oscillating behavior
is related to  a bifurcation of the solution to the linearized Usadel
equations.

Diffusive SNS junctions in a magnetic field have
been the subject of recent experimental studies,\cite{angers:08,chiodi:12}
and both the Fraunhofer regime and the purely decaying have been
observed. The oscillating-exponential regime predicted in our present paper
has not been yet identified in experiments. We propose that
it may be most conveniently observed
in nearly square junctions with $L_x \gtrsim L_y$. In such junctions,
the purely decaying regime at low fields crosses over to the
oscillating-exponential regime at higher (but not very high)
fields, see Fig.~\ref{fig:phase}.
A robust qualitative feature of this crossover (following from our
analysis and observed in the numerical studies of 
Ref.~\onlinecite{bergeret:07:08}) is the flux of the {\it first} 
reversal of the critical current being larger than the subsequent period
of current reversals [which tends to $\phi_0$ at large fields,
see Eq.~(\ref{current-field-wide})].

Such an experiment may, for example, be realized using SNS junctions
similar to those studied in Ref.~\onlinecite{chiodi:12}, but with
intermediate ratios $L_x/L_y$. If one considers a junction 
similar to the WAu-Sq sample of Ref.~\onlinecite{chiodi:12}, but with
$L_x=1.2~\mu$m and $L_y=1.0~\mu$m, then the crossover between the 
purely decaying and the oscillating-decaying phases occurs around 
$\xi_H \sim 1.2~\mu$m (the bifurcation point in Fig.~\ref{fig:effective}), 
which corresponds to the magnetic field $H \sim 14~$G. Using the Thouless energy 
$E_{\rm Th}=5.3~\mu$eV reported in Ref.~\onlinecite{chiodi:12} for
their WAu-Sq sample, we estimate the magnetic energy scale (\ref{EH})
in our example at this magnetic field to be $E_H \sim E_{\rm Th} \sim 60~$mK. 
Therefore, for such a junction, at $T\sim 60~$mK (the temperature used in 
Ref.~\onlinecite{chiodi:12}), we expect that our results (assuming
$T\ll E_H$) would hold qualitatively, but not quantitatively.
A quantitative agreement would improve at higher fields
(deeper in the damped-oscillating region) or at lower temperatures.

\begin{acknowledgments}
This work was supported by the Swiss National Foundation. We thank 
P.~Ostrovsky, M.~Skvortsov, and S.~Tollis for helpful discussions.
\end{acknowledgments}

\appendix

\section{Gauge choice and neglecting the magnetic penetration length}
\label{appendix:gauge}

We fix the gauge of the magnetic field in such a way that the phase of
the superconducting order parameter is constant in space in each
of the two leads. With this gauge choice, the vector potential in the
superconducting leads is proportional to the Meissner screening current.
Assuming that the field is uniform in the normal part of the junction
and exponentially decaying on the penetration length $\lambda$ inside
the superconductors (Meissner effect), we choose the gauge as follows:
\begin{equation}
\mathbf{A} = \begin{cases}
\lambda H e^{x/\lambda}\, \mathbf{e}_y\, , &  x<0\, , \\
\lambda H \left( 1- \frac{2x}{L_x}\right) \mathbf{e}_y  
-  y H \left( 1+ \frac{2\lambda}{L_x} \right)\mathbf{e}_x \, , \hspace{-3cm} & \\
 &  0 < x < L_x\, , \\
- \lambda H e^{-(x-L_x)/\lambda}\,  \mathbf{e}_y\, , &  x> L_x \, .
\end{cases}
\label{finite-lambda}
\end{equation}

In the main body of our paper, we take the limit $\lambda \to 0$: then the Meissner
screening current is also weak, and the vector potential (\ref{finite-lambda})
reduces to 
\begin{equation}
\mathbf{A} = \begin{cases}
-  y H \mathbf{e}_x \, , &  0 < x < L_x\, , \\
0\, ,  & x<0 \quad \text{or} \quad x>L_x\, .
\end{cases}
\label{zero-lambda}
\end{equation}
This approximation has been frequently used in earlier studies of SNS and SN systems
in a magnetic field.\cite{bergeret:07:08,galaiko-gogadze-antsygina,belzig:96}
It amounts to neglecting the $y$ component of the vector potential (\ref{finite-lambda}),
which is justified as long as $A_y \ll \phi_0 \xi_H^{-1}$. This condition may be rewritten
as $H\ll \phi_0/\lambda^2 \sim H_{c1}$ (the lower critical field). This estimate can 
also be applied to type I superconductors, where it suffices to assume that the field
is much lower than the critical field.

Note that the finite value of $\lambda$ may be taken into account in the
renormalization of the $x$ component of the vector potential (\ref{finite-lambda}).
In other words, the vector potential (\ref{zero-lambda}) used in the derivations
in the main body of the paper involves the effective field 
$H_{\rm eff}=H(L_x +2\lambda)/L_x$ instead of the actual field $H$. As a result,
the flux $\phi$ in all our formulas should be understood as the {\it total}
flux through the junction, including the field penetrating in the superconducting
leads.

\section{Estimating the magnitude of the critical current}
\label{appendix:current}

The estimate of the critical current $I_c^{(0)}$ in Eq.~(\ref{current3})
depends on whether the temperature $T$ is high or low (either one or many
Matsubara frequencies contribute) and on the boundary
conditions at the SN interface.

In the large-$L_x$ limit assumed in our derivation, the main dependence
on the Matsubara frequency comes from the exponent $\exp[-\kappa L_x]$
in Eqs.\ (\ref{current1}) and (\ref{current2}). Therefore the temperature
scale $T_*$ separating the high- and low-temperature limits may be
estimated from the condition
\begin{equation}
T_* \Re \frac{\partial \kappa}{\partial \omega} L_x \sim 1\, .
\end{equation}
In turn, $\partial \kappa / \partial \omega$ may be estimated by a variation
of Eq.~(\ref{zero-mode}):
\begin{equation}
\frac{\partial \kappa}{\partial \omega} = \frac{
\int_{-L_y/2}^{L_y/2} \psi^2(y) dy }{ E_H
\int_{-L_y/2}^{L_y/2} (\kappa - 2\pi i y) \psi^2(y) dy} \sim \frac{1}{E_H \langle \kappa - 2\pi i y \rangle}\, .
\end{equation}
The average $\langle \kappa - 2\pi i y \rangle$ is taken over the region
of $y$ where the mode $\psi(y)$ is located. In the narrow-junction regime $L_y \ll \xi_H$,
$\psi(y)$ is spread over the whole junction (Fig.~\ref{fig:zero-modes}), and
$\langle \kappa - 2\pi i y \rangle \sim \kappa \sim L_y$ [see Eq.~(\ref{narrow})].
In the wide-junction regime $L_y \gg \xi_H$, $\psi(y)$ is localized near the edge
of the junction (Fig.~\ref{fig:zero-modes}), and 
$\langle \kappa - 2\pi i y \rangle \sim \kappa_\mathrm{res} \sim 1$ 
[see Eq.~(\ref{kappa-res-1})]. Combining all these
considerations, we arrive at the estimate (\ref{T-star}) for $T_*$.

At $T \gg T_*$, only the lowest Matsubara frequency is relevant, and the
currents (\ref{current1}) and (\ref{current2}) may be estimated as
\begin{equation}
I_c^{(0)} \sim e\nu D L_z T \int_{-L_y/2}^{L_y/2} (\kappa - 2\pi i y) \psi^2(y) dy\, .
\label{Ic0-estimate}
\end{equation}
Assuming the rigid boundary conditions (\ref{rigid-bc}), we estimate
the overall amplitude of $\psi(y)$ to be of order one, and therefore
\begin{equation}
\int_{-L_y/2}^{L_y/2} (\kappa - 2\pi i y) \psi^2(y) dy \sim
\begin{cases}
(L_y/\xi_H)^2 \, , & L_y \ll \xi_H\, , \\
1 \, , & L_y \gg \xi_H\, .
\end{cases}
\end{equation}
This produces the high-temperature estimates in Eqs.\ (\ref{current-narrow-1}) 
and (\ref{current-wide-1}). For convenience, we express $I_c^{(0)}$ in terms
of the critical current
\begin{equation}
I_0 \sim e\nu D L_z E_\mathrm{Th} \frac{L_y}{L_x}
\end{equation}
at low temperature ($T\ll E_\mathrm{Th}$) through the same junction
at zero magnetic field.\cite{Zaikin-Zharkov:81,Dubos:01}

At $T \ll T_*$, one needs to sum over many Matsubara frequencies,
up to the frequencies of the order $T_*$. As a consequence, the
same estimate (\ref{Ic0-estimate}) applies, but with $T$ replaced by $T_*$.
This immediately produces the low-temperature estimates 
in Eqs.\ (\ref{current-narrow-1}) and (\ref{current-wide-1}).

\section{Fraunhofer interference pattern in wide and short junctions}
\label{appendix:Fraunhofer}

For completeness of our discussion, we remark that our model of Section
\ref{section:junction} produces Fraunhofer-like 
interference pattern\cite{josephson:64,barone:82}
in the limit $L_x \ll \xi_H$ and $L_y \gg \xi_H$. In such wide and short
junctions, the gradients along the $y$ direction in the Usadel equation
(\ref{usadel}) may be neglected, and the resulting system is equivalent
to an ensemble of Josephson junctions connected in parallel, with
the phase difference linearly depending on $y$. The gauge-invariant
phase difference across the junction is
\begin{equation}
\varphi(y) = 2\chi - \frac{2e}{\hbar} \int \mathbf{A}_x \, dx = 2\chi + \frac{2e}{\hbar} y H L_x
\end{equation}
(so that $2\chi$ equals the gauge-invariant phase difference 
in the middle of the junction $y=0$).
The critical current is then
\begin{multline}
I_c = L_z \max_\chi \int_{-L_y/2}^{L_y/2} j \left( 2\chi + \frac{2e}{\hbar} y H L_x \right)\, dy \\
= \frac{L_y L_z\phi_0}{2\pi\phi} \max_\chi \int_{-\pi\phi/\phi_0}^{\pi\phi/\phi_0} j(2\chi + \phi')\, d\phi' \, ,
\label{fraunhofer-1}
\end{multline}
where $j(\varphi)$ is the Josephson current density without magnetic field
at the phase difference $\varphi$ across the junction. At temperatures
$T \gg E_\mathrm{Th}$, the current-phase relation is sinusoidal,\cite{Zaikin-Zharkov:81,Dubos:01}
and the Fraunhofer pattern (\ref{fraunhofer-1}) takes the simplest form,
\begin{equation}
I_c \propto \frac{\sin(\pi\phi/\phi_0)}{\pi\phi/\phi_0} \, .
\label{fraunhofer-2}
\end{equation}
At lower temperature, the current-phase relation $j(\varphi)$ is
non-sinusoidal, and the sine function in Eq.~(\ref{fraunhofer-2}) is
replaced by a different, albeit qualitatively similar, periodic
dependence with the same period. We remark that this periodicity
is the same as in the decaying-oscillating regime (\ref{current-field-wide}).


\begin{thebibliography}{99}
\bibitem{note:Macdonald}
The Macdonald function
$K_{3/4}(z)$ does not have zeros on the principal sheet of the Riemann
surface, therefore the solution must have 
$\pi/2 <|\arg (-i \kappa_{\mathrm{res}})| < \pi$. We also impose the condition 
$\Re \kappa_{\mathrm{res}} > 0$, since the mode decays along the $x$ 
direction. Finally, under these constraints, we select the root with 
the smallest real part. Technically, the second sheet of the
Riemann surface may be accessed with the relation
$K_{3/4}(e^{i\pi} z) = e^{-3 \pi i/4} K_{3/4} (z) - i\pi I_{3/4} (z)$.
\end{thebibliography}
\end{document}